\documentclass[a4paper,onecolumn,9pt]{article}
\usepackage[latin1]{inputenc}
\usepackage[english]{babel}
\usepackage{epsfig}
\usepackage{latexsym}
\usepackage{amsmath}
\usepackage{a4wide}
\usepackage{amssymb}
\usepackage{graphicx}

\begin{document}

\begin{center}
{\LARGE \textbf{Magnetic fields from inflation?}}
\vspace{33pt}

  {{\bf  Vittoria Demozzi$^{1}$,  Viatcheslav Mukhanov$^{1,2}$ and Hector Rubinstein$^{3}$}}

    \vspace{20pt}

 {$^{1}$Arnold-Sommerfeld-Center for Theoretical Physics, Department f\"ur Physik,
Ludwig-Maximilians-Universit\"at M\"unchen,
Theresienstr. 37, D-80333, Munich, Germany}

\vspace{5pt}

{$^{2}$CCPP, Department of Physics, New York University,
4 Washington Place, New York, NY 10003}

\vspace{5pt}

{$^{3}$The Oskar Klein Centre for Cosmoparticle Physics, Department of Physics,
Stockholm University, AlbaNova University Center, SE - 106 91 Stockholm, Sweden}

\end{center}

\vspace{7pt}

\begin{center}
{\small vittoria.demozzi@physik.uni-muenchen.de; \hspace{3pt} viatcheslav.mukhanov@physik.uni-muenchen.de;\\
rub@fysik.su.se}
\end{center}

\vspace{20pt}

\date{}

\begin{abstract}
We consider the possibility of generation of the seeds of primordial magnetic
field on inflation and show that the effect of the back reaction of this field
can be very important. Assuming that back reaction does not spoil inflation
we find a rather strong restriction on the amplitude of the primordial seeds
which could be generated on inflation. Namely, this amplitude recalculated
to the present epoch cannot exceed $10^{-32}G$ in $Mpc$ scales. This field
seems to be too small to be amplified to the observable values by galactic
dynamo mechanism.
\end{abstract}

\vspace{20pt}

\section{Introduction}

Astronomical observations show that all celestial bodies carry magnetic
fields. From planets to interstellar medium, fields of varying strength and
extension have been measured. A particular interesting case are galaxies,
galaxy clusters and beyond, the intergalactic medium (IGM) and the Universe
at large. These fields are of order of a few micro Gauss and they extend
over kiloparsecs or more. Unfortunately their structure is not always
simple. Besides a constant component they have complex structure with varying symmetry, that shows that processing has taken
place since their appearance. 

The origin of magnetic fields is unknown and many scenarios have been
proposed to explain them. Until recently the most accepted idea for the
formation of large-scale magnetic fields was the exponentiation of a seed
field as suggested by Zeldovich and collaborators long ago. This seed
mechanism is known as galactic dynamo, the idea is the amplification of a
tiny field created early enough by differential rotation of the galaxies and
the subsequent generation of the galactic and cluster fields.\newline
However recent observational developments have cast serious doubts on this
possibility. In fact there are already many reasons to believe, tough this
is a possible mechanism in some cases, it cannot be universal \cite{bernet}.
Some of the reasons to think that seeding cannot be an answer are simple 
\cite{brandenburg}, \cite{rubinstein}. First, the very existence of high z galaxies with fields
comparable to the Milky Way is incompatible with the necessary number of
turns. Second, the narrowness of the distribution, most galaxies and
clusters have fields of a few micro Gauss, and this is not compatible with
the different number of rotations and the parameters involved in every
galaxy. Furthermore, magnetic fields seem to increase with redshift, though
the evidence is not overwhelming, the sample of Faraday rotations measured
is now consistent with an increase and the set includes tens of galaxies
showing this pattern. Finally, as pointed out by Dolgov, it is difficult to
create the fields in clusters since even the most efficient ejection from
point bodies in galaxies like supernovas would have difficulty creating
them. All put together seeding seems to be ruled out and moreover, even if
the galactic dynamo was effective, one should justify the presence of a seed
field which started the process.

This is why the mechanism responsible for the origin of large-scale magnetic
fields is looked in the Early Universe \cite{rubinstein}, \cite{dolgov}, \cite{dolgov2}.
In this paper we will consider the generation of large-scale magnetic fields
during inflation, which as it was noted by Turner and Widrow is a natural
candidate for doing this job \cite{turner}.\newline
It is known that in the Friedmann universe the conformal vacuum is preserved
if the theory is conformally invariant \cite{parker}. Classical
elctrodynamics is conformally invariant, so that photons should not be
produced in cosmological background. Thus the conformal invariance of the
electromagnetism must be broken to produce long wave magnetic fields via
excitation of the vacuum fluctuations.

Different mechanisms to break the conformal invariance of electromagnetic
field were proposed in ref. \cite{turner}. All of them are effectively
reduced either to the appearance of the effective mass or time dependent
coupling constant \cite{turner}, \cite{silk}, \cite{dolgov3}, \cite{tornkvist},
 \cite{ratra}, \cite{peebles}, \cite{lemoine}, \cite{gasperini}, \cite{bassett}, \cite{sasaki}, \cite{durrer}.

Here we will consider the broad class of models where conformal invariance
is broken during inflation and investigate the back reaction of the generated magnetic field on the
background. We show that this back reaction is very important and leads to
rather strong bounds on the maximal value of the strength of primordial
magnetic fields which seems not enough to explain the observed fields as a
result of amplification of these primordial seeds by dynamo mechanism.

\section{Models}

The action for the massless vector field is 
\begin{equation}
S=-\frac{1}{4}\int F_{\mu \nu }F^{\mu \nu }\sqrt{-g}d^{4}x=-\frac{1}{4}\int
F_{\mu \rho }F_{\nu \sigma }g^{\mu \nu }g^{\rho \sigma }\sqrt{-g}d^{4}x,
\label{1}
\end{equation}%
where $F_{\mu \nu }=\nabla _{\mu }A_{\nu }-\nabla _{\nu }A_{\mu }=\partial
_{\mu }A_{\nu }-\partial _{\nu }A_{\mu }$ is conformally invariant. It is
easy to see that under conformal transformation $g_{\mu \nu }\rightarrow
\Omega ^{2}g_{\mu \nu }$ the determinant transforms as $g\rightarrow \Omega
^{8}g$ and $g^{\mu \nu }\rightarrow \Omega ^{-2}g^{\mu \nu }.$ This is the
reason why in the Friedmann universe with the metric%
\begin{equation}
ds^{2}=a^{2}\left( \eta \right) \left( d\eta ^{2}-\delta
_{ik}dx^{i}dx^{k}\right)  \label{2}
\end{equation}%
the conformal vacuum is preserved. Therefore if we want to amplify quantum
fluctuations on inflation and thus explain the origin of primordial magnetic
fields we have to assume that either electromagnetic field is massive or its
effective coupling is time-dependent during inflation. Both of these options
are taken into account if we write the action in the form%
\begin{equation}
S=\int \left( -\frac{1}{4}I^{2}F_{\mu \nu }F^{\mu \nu }+M^{2}A_{\mu }A^{\mu
}\right) \sqrt{-g}d^{4}x.  \label{3}
\end{equation}%
Here $I(t)=I(\phi (t),...)$, where $\phi $ can be the inflaton, dilaton or
some other scalar field and the dots can be anything, for instance,
invariants of the curvature (see \cite{ratra}, \cite{sasaki}, \cite{gasperini}, \cite{gasperini2}). The appearance of time dependence of the
coefficient in front of $F^{2}$ term is naturally interpreted as
time-dependent coupling constant of the vector field. In fact if we write
the Lagrangian density of the vector field coupled with a charged fermion in
the standard form as 
\begin{equation}
L=-\frac{1}{4}F_{\mu \nu }F^{\mu \nu }+i\bar{\psi}\gamma ^{\mu }(\partial
_{\mu }+igA_{\mu })\psi ,  \label{3a}
\end{equation}%
where $g$ is the coupling constant, then after rescaling the vector
potential by the coupling constant $A_{\mu }\rightarrow gA_{\mu }$ we bring
this Lagrangian to the form%
\begin{equation}
L=-\frac{1}{4g^{2}}F_{\mu \nu }F^{\mu \nu }+i\bar{\psi}\gamma ^{\mu
}(\partial _{\mu }+iA_{\mu })\psi ,  \label{3b}
\end{equation}%
which is \textquotedblleft ready\textquotedblright\ for introducing a
time-dependent coupling constant. Note that $I$ is an inverse coupling
constant and small values of $I$ correspond to a large coupling constant $g$, which in turn would mean that we are in uncontrollable strong coupling
regime. Only if $I$ is large we can trust the theory. For our purposes we do
not need to specify in more details the origin of the time dependence of $I$
here. Note that the time-dependent effective coupling leaves the Lagrangian
to be $U\left( 1\right) $ gauge-invariant.

The mass term introduced by \textquotedblleft hand\textquotedblright\ spoils
gauge invariance. Only when it is generated via Higgs mechanism the gauge
invariance is preserved. On the other hand as it was noticed already in \cite%
{turner} large enough magnetic fields can be obtained only if $M^{2}$ is
negative during inflation. However, to generate negative mass squared term
via Higgs mechanism one needs a ghost scalar field with negative kinetic
energy \cite{dvali}, \cite{himmetoglu}. As it is well known ghosts lead to
catastrophic instabilities and therefore we will not exploit this
possibility any further here. Instead we introduce the effective negative
mass square terms considering the non-minimal coupling of the vector field
to gravity, so that,%
\begin{equation}
M^{2}=m^{2}+\xi R,  \label{4}
\end{equation}%
where for generality we also keep \textquotedblleft hard\textquotedblright\
mass term $m^{2}$ assuming that it is positive.

Let us now rewrite the action (\ref{3}) in terms of the vector potential $%
A_{\alpha }=\left( A_{0},A_{i}\right) .$ It is convenient to decompose the
spatial part of the vector potential in terms of its transverse and
longitudinal components $A_{i}=A_{i}^{T}+\partial _{i}\chi ,$ where $%
\partial _{i}A_{i}^{T}=0$ (we will be assuming summation over repeated
indices irrespective of their position). In the homogeneous flat universe
with metric (\ref{2}), the action (\ref{3}) then becomes%
\begin{eqnarray}
S &=&\frac{1}{2}\int \left[ I^{2}\left( A_{i}^{T\prime }A_{i}^{T\prime
}+A_{i}^{T}\Delta A_{i}^{T}+2A_{0}\Delta \chi ^{\prime }-A_{0}\Delta
A_{0}-\chi ^{\prime }\Delta \chi ^{\prime }\right) \right.  \notag \\
&&+\left. M^{2}a^{2}\left( A_{0}^{2}+\chi \Delta \chi
-A_{i}^{T}A_{i}^{T}\right) \right] d^{4}x,  \label{5}
\end{eqnarray}%
where prime denotes derivative with respect to the conformal time $\eta .$
We will consider different cases separately.

\subsection{Time dependent coupling}

Let us first consider the case when $M^{2}=0$ and $I=I\left( t\right) .$
Then the variation of the action (\ref{5}) with respect to $A_{0}$ gives $%
A_{0}=\chi ^{\prime },$ and the action simplifies to%
\begin{equation}
S=\frac{1}{2}\int I^{2}\left( A_{i}^{T\prime }A_{i}^{T\prime
}+A_{i}^{T}\Delta A_{i}^{T}\right) d^{4}x.  \label{6}
\end{equation}%
Substituting the expansion 
\begin{equation}
A_{i}^{T}(\mathbf{x,}\eta )=\sum_{\sigma =1,2}\int A_{\mathbf{k}}^{(\sigma
)}(\eta )\varepsilon _{i}^{(\sigma )}(\mathbf{k})e^{i\mathbf{k}\cdot \mathbf{%
x}}\frac{d^{3}k}{(2\pi )^{3/2}},  \label{7}
\end{equation}%
where $\varepsilon _{i}^{(\sigma )}(\mathbf{k})$, $\sigma =1,2$ are two
orthogonal polarization vectors, into (\ref{6}), we obtain%
\begin{equation}
S=\frac{1}{2}\sum_{\sigma =1,2}\int I^{2}\varepsilon _{i}^{(\sigma )}(%
\mathbf{k})\varepsilon _{i}^{(\sigma )}(-\mathbf{k})\left( A_{\mathbf{k}%
}^{(\sigma )\prime }A_{-\mathbf{k}}^{(\sigma )\prime }-k^{2}A_{\mathbf{k}%
}^{(\sigma )}A_{-\mathbf{k}}^{(\sigma )}\right) d\eta d^{3}k.  \label{8}
\end{equation}%
Rewritten in terms of the new variable%
\begin{equation}
v_{\mathbf{k}}^{(\sigma )}=\sqrt{\varepsilon _{i}^{(\sigma )}\varepsilon
_{i}^{(\sigma )}}IA_{\mathbf{k}}^{(\sigma )},  \label{9}
\end{equation}%
this action becomes 
\begin{equation}
S=\frac{1}{2}\sum_{\sigma =1,2}\int \left( v_{\mathbf{k}}^{(\sigma )\prime
}v_{-\mathbf{k}}^{(\sigma )\prime }-\left( k^{2}-\frac{I^{\prime \prime }}{I}%
\right) v_{\mathbf{k}}^{(\sigma )}v_{-\mathbf{k}}^{(\sigma )}\right) d\eta
d^{3}k.  \label{10}
\end{equation}%
It describes two real scalar fields with time-dependent effective masses in
terms of their Fourier components.

We are interested in the correlation functions of the transverse components
of the vector potential and magnetic field assuming that initially the field
was in its vacuum state. The quantization of the fields with action (\ref{10}%
) is standard and we will simply summarize here the results referring the
reader to \cite{mukhanov-book1}, \cite{mukhanov-book} for the details.
Taking into account (\ref{9}) and (\ref{7}), we immediately find the
correlation function%
\begin{equation}
<0|\hat{A}_{i}^{T}(\eta ,\mathbf{x})\hat{A}^{Ti}(\eta ,\mathbf{y})|0>=-\frac{%
1}{4\pi ^{2}a^{2}I^{2}}\sum_{\sigma =1,2}\int |v_{\mathbf{k}}^{(\sigma
)}(\eta )|^{2}k^{3}\frac{\sin {k|\mathbf{x}-\mathbf{y}|}}{k|{\mathbf{x}-%
\mathbf{y}}|}\frac{dk}{k},  \label{11}
\end{equation}%
where $v_{k}^{(\sigma )}(\eta )$ satisfy the equations 
\begin{equation}
v_{\mathbf{k}}^{(\sigma )\prime \prime }+\omega ^{2}\left( \eta \right) v_{%
\mathbf{k}}^{(\sigma )}=0,\text{ \ \ }\omega ^{2}\left( \eta \right) \equiv
\left( k^{2}-\frac{I^{\prime \prime }}{I}\right) ,  \label{12}
\end{equation}%
which immediately follow from action (\ref{10}). The initial conditions for
these equations corresponding to the initial vacuum state at $\eta _{i}$ are 
\begin{equation}
v_{\mathbf{k}}^{(\sigma )}\left( \eta _{i}\right) =\frac{1}{\sqrt{\omega
\left( \eta _{i}\right) }},\text{ \ \ }v_{\mathbf{k}}^{(\sigma )\prime
}\left( \eta _{i}\right) =i\sqrt{\omega \left( \eta _{i}\right) }.
\label{13}
\end{equation}%
These initial conditions make sense only if $\omega ^{2}>0.$ Anyway we will
need them only for the short-wavelength modes for which $\omega ^{2}\simeq
k^{2}.$ The power spectrum characterizing the typical amplitude squared of
the invariant magnitude of the vector potential, $A=\sqrt{-A_{i}A^{i}},$ in
the appropriate comoving scale $\lambda =2\pi /k$ is 
\begin{equation}
\delta _{A}^{2}\left( k,\eta \right) =\sum_{\sigma =1,2}\frac{|v_{\mathbf{k}%
}^{(\sigma )}(\eta )|^{2}k^{3}}{4\pi ^{2}a^{2}I^{2}}.  \label{14}
\end{equation}%
Taking into account that the magnitude of the magnetic field is%
\begin{equation}
B^{2}=-B_{i}B^{i}=\frac{1}{2a^{4}}F_{ik}F_{ik}=\frac{1}{a^{4}}\left(
\partial _{i}A_{k}\partial _{i}A_{k}-\partial _{k}A_{i}\partial
_{i}A_{k}\right),  \label{15}
\end{equation}%
we obtain for the power spectrum of the magnetic field 
\begin{equation}
\delta _{B}^{2}\left( k,\eta \right) =\delta _{A}^{2}\left( k,\eta \right) 
\frac{k^{2}}{a^{2}}=\sum_{\sigma =1,2}\frac{|v_{\mathbf{k}}^{(\sigma )}(\eta
)|^{2}k^{5}}{4\pi ^{2}a^{4}I^{2}},  \label{16}
\end{equation}%
that is, its amplitude decays faster by an extra power of the scale compared
to the amplitude of the vector potential. For example, a flat spectrum for
magnetic field ($\delta _{B}\left( k\right) =const$) corresponds to the
linearly growing towards large scales spectrum for the vector potential,
that is, $\delta _{A}\left( k,\eta \right) \propto k^{-1}.$

We will need to control the back reaction of the generated electromagnetic
field on the background. With this purpose let us calculate the expectation
value of the energy density equal to $T_{0}^{0}$ component of the
energy-momentum tensor:%
\begin{equation}
T_{0}^{0}=I^{2}\left( \frac{1}{4}F_{\alpha \beta }F^{\alpha \beta
}-F_{0\alpha }F^{0\alpha }\right) =\frac{I^{2}}{2a^{4}}\left( A_{i}^{T\prime
}A_{i}^{T\prime }+\partial _{i}A_{k}^{T}\partial _{i}A_{k}^{T}\right) .
\label{17a}
\end{equation}%
Taking into account (\ref{7}) and (\ref{9}) we obtain%
\begin{equation}
\varepsilon _{EM}=<0|\hat{T}_{0}^{0}|0>=\frac{1}{8\pi ^{2}a^{4}}\sum_{\sigma
=1,2}\int \left[ |v_{k}^{(\sigma )\prime }(\eta )|^{2}-\frac{I^{\prime }}{I}%
|v_{k}^{(\sigma )}(\eta )|^{2\prime }+\left( \frac{I^{\prime 2}}{I^{2}}%
+k^{2}\right) |v_{k}^{(\sigma )}(\eta )|^{2}\right] k^{3}\frac{dk}{k}.
\label{17b}
\end{equation}

Let us assume that the function $I$ depends on time during inflation and
find the resulting spectrum of the magnetic field at the end of inflation.
For short waves with $k\left\vert \eta \right\vert $ $\gg 1$ we can neglect $%
I^{\prime \prime }/I$ compared to $k^{2}$ in (\ref{12}) and the solution of
this equation with vacuum initial conditions (\ref{13}) then becomes 
\begin{equation}
v_{k}^{(\sigma )}\left( \eta \right) \simeq \frac{1}{\sqrt{k}}e^{ik\left(
\eta -\eta _{i}\right) }.  \label{17}
\end{equation}%
Because $\left\vert \eta \right\vert $ decreases during inflation at some
moment $\left\vert \eta _{k}\right\vert \simeq 1/k$ the physical scale of
the wave with comoving wavenumber $k$ begins to exceed the curvature scale
and taking into account that $k^{2}\ll I^{\prime \prime }/I$ we can write
the general longwave solution of (\ref{12}) as 
\begin{equation}
v_{k}^{(\sigma )}\left( \eta \right) \simeq C_{1}I+C_{2}I\int \frac{d\eta }{%
I^{2}},  \label{18}
\end{equation}%
where $C_{1}$ and $C_{2}$ are the constants of integration which have to be
fixed by matching this solution to (\ref{17}) at $\left\vert \eta
_{k}\right\vert \simeq 1/k.$ Let us assume that $I$ is a power-law
function of the scale factor during inflation%
\begin{equation}
I=I_{f}\left( \frac{a}{a_{f}}\right) ^{n},  \label{19}
\end{equation}%
where $a_{f}$ is the scale factor at the end of inflation. Taking into
account that 
\begin{equation*}
d\eta =\frac{da}{Ha^{2}},
\end{equation*}%
and the Hubble constant $H$ does not change significantly during inflation
we obtain from (\ref{18})%
\begin{equation}
v_{k}^{(\sigma )}\left( \eta \right) \simeq C_{1}a^{n}+C_{2}a^{-n-1}.
\label{20}
\end{equation}

\subsubsection{Strong coupling case}

In case $n>-1/2$ the first mode dominates and, matching solutions (\ref{17})
and (\ref{20}) at $\left\vert \eta _{k}\right\vert \simeq 1/k,$ we find%
\begin{equation}
v_{k}^{(\sigma )}\left( \eta \right) \simeq \frac{1}{\sqrt{k}}\left( \frac{a%
}{a_{k}}\right) ^{n}\simeq \frac{1}{\sqrt{k}}\left( \frac{H_{I}a}{k}\right)
^{n},  \label{21}
\end{equation}%
where we have taken into account that at the moment $\eta _{k},$ when the
corresponding wave crosses the Hubble scale, the scale factor is $a_{k}\simeq
k/H_{I}$. Substituting (\ref{21}) into (\ref{16}) we obtain at the end of
inflation 
\begin{equation}
\delta _{B}\left( \lambda _{ph},\eta _{f}\right) \simeq \frac{H_{I}^{2}}{%
\sqrt{2}\pi I_{f}}\left( \frac{\lambda _{ph}}{H_{I}^{-1}}\right) ^{n-2},
\label{22}
\end{equation}%
where $\lambda _{ph}=a_{f}/k$ is the physical wavelength and $H_{I}$ is the
Hubble constant on inflation. This formula is valid for $H_{I}^{-1}\left(
a_{f}/a_{i}\right) >\lambda _{ph}>H_{I}^{-1}$, where $a_{i}$ is the value of
the scale factor at the beginning of inflation. If $n=2$ the spectrum of the
magnetic field is flat. For $H_{I}^{2}\simeq 10^{-12}$ (in Planck units),
required by primordial inhomogeneities \cite{mukhanov-book1}, and $%
I_{f}\simeq O\left( 1\right) ,$ the amplitude of the field is the same in
all scales and it is equal to $\delta _{B}\simeq 10^{-12}$ Planck units or $%
\sim 10^{46}$ $G$ immediately after inflation. Later on the magnetic field
is frozen and decays inversely proportional to the scale factor squared. To
estimate how much the scale factor increases after inflation we can use the
entropy conservation law. Assuming that inflation is followed but the dust
dominated stage we obtain%
\begin{equation}
\frac{a_{0}}{a_{f}}\simeq g^{1/12}\frac{H_{I}^{1/2}}{T_{0}}\left( \frac{a_{R}%
}{a_{f}}\right) ^{1/4},  \label{23}
\end{equation}%
where $g$ is the number of relativistic degrees of freedom of those
particles which later on transfer their entropy to the photons, $T_{0}$ is
the temperature of the background radiation today and $a_{R}$ the scale
factor at the moment of reheating. The lower bound on this ratio is obtained
assuming that reheating happens immediately after inflation. In this case
for $H_{I}\simeq 10^{-6}$ we have $a_{0}/a_{f}\simeq 10^{29}$ and
correspondingly the strength of the generated magnetic field cannot exceed $%
10^{-12}$ $G.$

Let us calculate the energy density of the generated magnetic field. The
main contribution to the energy density comes from the scales exceeding $%
H_{I}^{-1}$ because the contribution from the subhorizon scales is
renormalized in the leading order. In the case when the dominant mode $%
v\propto I $ and $A^{T}\propto v/I\propto const$ the time derivatives of the
vector potential in (\ref{17a}) contribute only in subleading $k^{2}$ order
and their contribution is comparable to the contribution of the magnetic
field itself given by the last term in (\ref{17b}). Thus we obtain%
\begin{equation}
\varepsilon _{EM}=\frac{O\left( 1\right) }{a^{4}}%
\int_{H_{I}a_{i}}^{H_{I}a}|v_{k}(\eta )|^{2}k^{4}dk,  \label{24}
\end{equation}%
where $a_{i}$ is the value of the scale factor at the beginning of
inflation. Substituting (\ref{21}) into (\ref{24}) we find that at the end
of inflation when $a=a_{f}$%
\begin{equation}
\varepsilon _{EM}=O\left( 1\right) H_{I}^{4}\times \left\{ 
\begin{array}{ccc}
\frac{1}{2-n}, &  & n<2, \\ 
\ln \left( \frac{a_{f}}{a_{i}}\right) , & \text{for} & n=2, \\ 
\frac{1}{n-2}\left( \frac{a_{f}}{a_{i}}\right) ^{2\left( n-2\right) }, &  & 
n>2.%
\end{array}%
\right.  \label{25}
\end{equation}%
We see that the magnetic field energy can be comparable with the energy
density of the background only for $n\geq 2.$ Requiring that inflation
should last at least 75 e-folds we obtain that the contribution of the
magnetic field energy density does not spoil inflation, that is, $%
\varepsilon _{EM}$ is smaller that $H_{I}^{2}$ until the end of inflation,
only if $n-2<0.2.$ Thus, we can have slightly growing toward large scales
spectrum of the magnetic field. In particular, for $n\simeq 2.2$ the
amplitude of the magnetic field in $Mpc$ scales can be larger by a factor $%
10^{5}$ compared to the considered above case of the flat spectrum, that is, 
$\delta _{B}$ $\simeq 10^{-7}$ $G$ today. This is the greatest amplitude of
the primordial magnetic field which we can obtain in the considered above
case. Note that the theory where $I$ grows with the scale factor corresponds
to the case when the effective coupling constant, which is inversely
proportional to $I$, is incredibly large at the beginning of inflation and
becomes of the order of one at the end of inflation. Hence at the beginning
we are in strongly coupled regime where such theory is not trustable at all.

The case considered above is the only one in which we can generate strong
enough fields on inflation. Let us show that in all other cases there is
very strong bound on the possible value of the generated field due to the
back reaction of this field on the background.

\subsubsection{Weak coupling case}

For $n<-1/2$ the second term in (\ref{20}) dominates and 
\begin{equation}
v_{k}\left( \eta \right) \propto a^{-n-1}.  \label{27}
\end{equation}%
In this case the result follows immediately by substituting in the formulae
(\ref{21}) and (\ref{22}) $-n-1$ instead of $n,$ so that%
\begin{equation}
v_{k}^{(\sigma )}\left( \eta \right) \simeq \frac{1}{\sqrt{k}}\left( \frac{a%
}{a_{k}}\right) ^{-n-1}\simeq \frac{1}{\sqrt{k}}\left( \frac{H_{I}a}{k}%
\right) ^{-n-1},  \label{28}
\end{equation}%
and%
\begin{equation}
\delta _{B}\left( \lambda _{ph},\eta _{f}\right) \simeq \frac{H_{I}^{2}}{%
\sqrt{2}\pi I_{f}}\left( \frac{\lambda _{ph}}{H_{I}^{-1}}\right) ^{-n-3}.
\label{29}
\end{equation}%
Thus the spectrum of the magnetic field is flat for $n=-3.$ This case
corresponds to the coupling constant growing as $I^{-1}\propto a^{3},$ that
is, it changes from extremely small values at the beginning of inflation to
values of order of unity at the end of inflation. Thus the theory is
trustable everywhere. However here the back reaction of the field is very
large because $A\propto v/I\propto a^{-2n-1}$ changes very fast and the main
contribution to the energy density comes from the time derivative of the
vector potential in (\ref{17a}), that is, from the electric field.
Substituting (\ref{28}) in (\ref{17b}) we obtain that at the end of inflation%
\begin{equation}
\varepsilon _{EM}\simeq \frac{4n^{2}+4n+1}{8\pi ^{2}}H_{I}^{4}\times \left\{ 
\begin{array}{ccc}
\frac{1}{n+2}, &  & n>-2, \\ 
\ln \left( \frac{a_{f}}{a_{i}}\right) , & \text{for} & n=-2, \\ 
-\frac{1}{n+2}\left( \frac{a_{f}}{a_{i}}\right) ^{-2\left( n+2\right) }, & 
& n<-2.%
\end{array}%
\right.  \label{30}
\end{equation}%
Requiring that inflation should last at least 75 e-folds, we find that $%
\varepsilon _{EM}<H_{I}^{2}$ at the end of inflation only if $n>-2.2.$
Thus the flat spectrum for magnetic field cannot be generated during
inflation because in this case the back reaction of the electromagnetic
field would spoil inflation too early. In the most favorable admissible case 
$n\simeq -2.2,$ the amplitude of the magnetic field decays as $\delta
_{B}\propto \lambda _{ph}^{-0.8}$ and its value cannot exceed $10^{-32}$ $G$
in $Mpc$ scales today. Thus in this model with weak coupling constant during
inflation one cannot explain the origin of the primordial magnetic field.

\subsection{Massive field}

Now we set $I=1$ and consider the case when magnetic fields are generated by
the mass term in the action. Variation of action (\ref{5}) with respect to $%
A_{0}$ gives%
\begin{equation}
\Delta \chi ^{\prime }-\Delta A_{0}+M^{2}a^{2}A_{0}=0.  \label{31}
\end{equation}%
Taking the Fourier transform 
\begin{equation}
\chi (\mathbf{x,}\eta )=\int \chi _{\mathbf{k}}(\eta )e^{i\mathbf{k}\cdot 
\mathbf{x}}\frac{d^{3}k}{(2\pi )^{3/2}},\text{ \ }A_{0}(\mathbf{x,}\eta
)=\int A_{0\mathbf{k}}(\eta )e^{i\mathbf{k}\cdot \mathbf{x}}\frac{d^{3}k}{%
(2\pi )^{3/2}},  \label{33}
\end{equation}%
we obtain from here 
\begin{equation}
A_{0\mathbf{k}}=\frac{k^{2}}{k^{2}+M^{2}a^{2}}\chi _{\mathbf{k}}^{\prime
}\equiv F_{k}\chi _{\mathbf{k}}^{\prime }.  \label{34}
\end{equation}%
Substituting into action (\ref{5}) the expansions (\ref{7}), (\ref{33}) and
using (\ref{34}) to express $A_{0\mathbf{k}}$ in terms $\chi _{\mathbf{k}%
}^{\prime }$ we obtain%
\begin{eqnarray}
S &=&\frac{1}{2}\sum_{\sigma =1,2}\int \left( v_{\mathbf{k}}^{(\sigma
)\prime }v_{-\mathbf{k}}^{(\sigma )\prime }-\left( k^{2}+M^{2}a^{2}\right)
v_{\mathbf{k}}^{(\sigma )}v_{-\mathbf{k}}^{(\sigma )}\right) d\eta d^{3}k 
\notag \\
&&+\frac{1}{2}\int sign\left( 1-F_{k}\right) \left( \bar{\chi}_{\mathbf{k}%
}^{\prime }\bar{\chi}_{-\mathbf{k}}^{\prime }-\left( k^{2}+M^{2}a^{2}-\tfrac{%
\sqrt{\left\vert 1-F_{k}\right\vert }^{^{\prime \prime }}}{\sqrt{\left\vert
1-F_{k}\right\vert }}\right) \bar{\chi}_{\mathbf{k}}\bar{\chi}_{-\mathbf{k}%
}\right) d\eta d^{3}k,  \label{35}
\end{eqnarray}%
where $v_{\mathbf{k}}^{(\sigma )}$ is defined in (\ref{9}) ($I=1$)$,$ and 
\begin{equation}
\bar{\chi}_{\mathbf{k}}=k\sqrt{\left\vert 1-F_{k}\right\vert }\chi _{\mathbf{%
k}}.  \label{36}
\end{equation}%
Thus we see that in the case of massive field the longitudinal degree of
freedom $\chi $ becomes dynamical. In the case of positive mass squared $%
F_{k}$ is always smaller than unity and therefore the sign in front of the
longitudinal part of the action is positive. However, if $M^{2}$ is negative
then $1-F_{k}$ is negative for high momentum modes with $k^{2}>M^{2}a^{2}$
and these modes have negative kinetic energy. The low momentum modes with $%
k^{2}<M^{2}a^{2}$ have positive kinetic energy because $F_{k}$ is negative
for them. Thus, introducing a tachyonic mass for the vector field in a
\textquotedblleft hard\textquotedblright\ way seems to lead inevitably to
the appearance of ghost for high momentum longitudinal modes \cite{dvali}.
Therefore if we want to avoid catastrophic instabilities related with ghost
fields we have to consider tachyonic vector field only as a low energy
effective field theory description of some unknown yet theory with
\textquotedblleft safe\textquotedblright\ ultraviolet completion. On the
other hand if negative effective mass appears as interaction with the
curvature, $M^{2}=\xi R,$ then the field is massless on scales smaller then
the typical distance between particles inducing the average curvature and
thus there is a natural ultraviolet cutoff in the theory. Note that this
argument is not directly applicable in the presence of the cosmological
constant. Let us assume that the problem of ghosts can be somehow solved and
proceed with the calculation of the magnetic field from inflation in the
theory with $M^{2}=m^{2}+\xi R.$ In the case $m=0$ the photon mass is $%
m_{\gamma }\sim R^{1/2}$, where $R^{1/2}\sim H$. Today it would be $%
m_{\gamma }=H_{today}\sim 10^{-33}eV$, well below the available experimental
limits on the photon mass. The breaking of charge conservation also
manifests itself only on scales of the horizon or larger ($\geq H^{-1}\sim
10^{28}cm$) and hence has no observable consequences.

The equations of motion for transverse and longitudinal modes follow
immediately from the action (\ref{35}):%
\begin{equation}
v_{\mathbf{k}}^{(\sigma )\prime \prime }+\left( k^{2}+M^{2}a^{2}\right) v_{%
\mathbf{k}}^{(\sigma )}=0,\text{ \ }  \label{37}
\end{equation}%
and 
\begin{equation}
\bar{\chi}_{\mathbf{k}}^{\prime \prime }+\left( k^{2}+M^{2}a^{2}-\tfrac{%
\sqrt{\left\vert 1-F_{k}\right\vert }^{^{\prime \prime }}}{\sqrt{\left\vert
1-F_{k}\right\vert }}\right) \bar{\chi}_{\mathbf{k}}=0.  \label{38}
\end{equation}%
Let us consider de Sitter universe where%
\begin{equation*}
a=-\frac{1}{H_{I}\eta }.
\end{equation*}%
Taking into account that $R=-12H_{I}^{2}$, for $m^{2}=0$ equation (\ref{37})
becomes%
\begin{equation}
v_{\mathbf{k}}^{(\sigma )\prime \prime }+\left( k^{2}-\frac{12\xi }{\eta ^{2}%
}\right) v_{\mathbf{k}}^{(\sigma )}=0.  \label{39}
\end{equation}%
For short waves with $k\left\vert \eta \right\vert $ $\gg 1$ the solution of
this equation corresponding to vacuum initial conditions is 
\begin{equation}
v_{k}^{(\sigma )}\left( \eta \right) \simeq \frac{1}{\sqrt{k}}e^{ik\left(
\eta -\eta _{i}\right) }.  \label{40}
\end{equation}%
For $k\left\vert \eta \right\vert $ $\ll 1$ we can neglect the $k^{2}$ term
in (\ref{39}) and the dominating longwavelength solution of this equation is%
\begin{equation}
v_{k}^{(\sigma )}\left( \eta \right) \simeq \frac{1}{\sqrt{k}}\left( \frac{%
H_{I}a}{k}\right) ^{n},\text{ \ \ }n=\frac{1}{2}\left( \sqrt{1+48\xi }%
-1\right) ,  \label{41}
\end{equation}%
where we use the matching conditions at $\left\vert \eta _{k}\right\vert
\simeq 1/k$ \ to fix the constant of integration. Since here the
calculations are very similar to those in the previous section we can
immediately write the result for the magnetic field%
\begin{equation}
\delta _{B}\left( \lambda _{ph},\eta _{f}\right) \simeq O\left( 1\right)
H_{I}^{2}\left( \frac{\lambda _{ph}}{H_{I}^{-1}}\right) ^{n-2}.  \label{42}
\end{equation}%
For $\xi =1/6$ we have $n=1$ and the spectrum linearly decays with the
scale. In this case its value today is about $10^{-37}$ $G$ in $Mpc$ scales.
The flat spectrum is obtained for $\xi =1/2.$ However, to find out whether
this case is possible we have to verify that the back reaction of the
magnetic field will not spoil inflation too early. In the energy density
also contributes the longitudinal mode and to determine its contribution we
will need a longwavelength solution for $\bar{\chi}_{\mathbf{k}}.$ It is
easy to check that the term which is different in the equations (\ref{37})
and (\ref{38}) can be neglected for both shortwave and longwave solutions
and hence%
\begin{equation}
\bar{\chi}_{k}\left( \eta \right) \simeq \frac{1}{\sqrt{k}}\left( \frac{%
H_{I}a}{k}\right) ^{n},\text{ \ \ }n=\frac{1}{2}\left( \sqrt{1+48\xi }%
-1\right) .  \label{43}
\end{equation}%
Variation of action (\ref{3}), where $I=1$ and $M^{2}=m^{2}+\xi R,$ with
respect to the metric gives 
\begin{eqnarray}
T_{\mu }^{\rho } &=&\frac{1}{4}\delta _{\mu }^{\rho }F_{\alpha \beta
}F^{\alpha \beta }-F^{\rho \beta }F_{\mu \beta }-\frac{1}{2}\delta _{\mu
}^{\rho }(m^{2}+\xi R)A_{\alpha }A^{\alpha }  \notag \\
&+&(m^{2}+\xi R)A_{\mu }A^{\rho }+\xi R_{\mu }^{\rho }A_{\alpha }A^{\alpha
}+\xi \lbrack \delta _{\mu }^{\rho }\nabla ^{\alpha }\nabla _{\alpha
}(A^{\beta }A_{\beta })-\nabla _{\mu }\nabla ^{\rho }(A^{\beta }A_{\beta })].
\label{45}
\end{eqnarray}%
As a result of straightforward but rather lengthy calculations we obtain%
\begin{equation*}
<0|\hat{T}_{0}^{0}|0>=\varepsilon _{T}+\varepsilon _{L},
\end{equation*}%
where%
\begin{equation}
\varepsilon _{T}=\frac{1}{8\pi ^{2}a^{4}}\sum_{\sigma =1,2}\int \left[
|v_{k}^{(\sigma )\prime }|^{2}-6\xi aH|v_{k}^{(\sigma )}|^{2\prime }+\left(
k^{2}+m^{2}a^{2}+6\xi H^{2}a^{2}\right) |v_{k}^{(\sigma )}|^{2}\right] k^{3}%
\frac{dk}{k}  \label{46}
\end{equation}%
is the contribution of the transverse modes, $H=a^{\prime }/a^{2}$ is the
Hubble constant. The contribution of the longitudinal mode is given by%
\begin{equation}
\varepsilon _{L}=\frac{1}{8\pi ^{2}a^{4}}\int \left( 1-F\right) \left[
\left( 1-6\xi bF\right) |\tilde{\chi}_{k}^{\prime }|^{2}-6\xi aH\left( \frac{%
1+F}{1-F}\right) |\tilde{\chi}_{k}|^{2\prime }+\left( \frac{m^{2}a^{2}+6\xi
H^{2}a^{2}}{1-F}\right) |\tilde{\chi}_{k}|^{2}\right] k^{3}\frac{dk}{k},
\label{47}
\end{equation}%
where 
\begin{equation}
\tilde{\chi}=\bar{\chi}/\sqrt{\left\vert 1-F_{k}\right\vert },\text{ \ \ \ \ 
}b=\frac{\dot{H}+7H^{2}+4H\frac{\dot{M}}{M}}{M^{2}}.  \label{48}
\end{equation}%
For the longwave modes with $k^{2}\ll \left\vert M^{2}a^{2}\right\vert $ we
have $F_{k}\ll 1,$ $\tilde{\chi}\simeq \bar{\chi}$ and their contribution to
the total energy density is the same as the contribution from the transverse
mode. It is interesting to note that the longitudinal mode is the ghost in
de Sitter background. However, in Friedmann universe filled by matter with
positive pressure it is not ghost in spite of the fact that the effective
mass squared is negative.

Substituting (\ref{41}) into (\ref{46}) we find that in the leading order
the contribution of the longwave modes into the energy density in the case $%
m=0$ is 
\begin{equation}
\varepsilon _{L}\simeq O\left( 1\right) \frac{H_{I}^{2}}{a^{2}}\left(
n^{2}-12n\xi +6\xi \right) \int_{H_{I}a_{i}}^{H_{I}a}|v_{k}(\eta
)|^{2}k^{2}dk,  \label{49}
\end{equation}%
and calculating the integral we obtain%
\begin{equation}
\varepsilon _{L}\simeq O\left( 1\right) H_{I}^{4}\left( n^{2}-12n\xi +6\xi
\right) \left\{ 
\begin{array}{ccc}
\frac{1}{1-n}, & \text{for} & n<1, \\ 
\frac{1}{n-1}\left( \frac{a}{a_{i}}\right) ^{2\left( n-1\right) }, & \text{for%
} & n>1.%
\end{array}%
\right.  \label{50}
\end{equation}%
In the case $\xi =1/6$ and when $n=1$ the contribution is canceled in the
leading order and $k^{2}$ terms give a contribution of the order of $%
H_{I}^{4},$ that is the same as for $n<1.$ However, for $\xi >1/6,$ and
correspondingly $n>1,$ the energy density of the longwavelengh
electromagnetic waves grows with time rather fast. It is negative and
therefore when it becomes of order $H_{I}^{2}$ inflation is over. Requiring
that inflation should last at least 75 e-fold we find that the contribution
of electromagnetic field does not spoil inflation only if $n-1<0.2.$ Thus,
in the most favorable case of $n\simeq 1.2,$ the amplitude of the magnetic
field decays as $\delta _{B}\propto \lambda _{ph}^{-0.8}$ and its value does
not exceed $10^{-32}$ $G$ in $Mpc$ scales today.

\section{Conclusions}

In this paper we have studied the generation of large-scale magnetic fields
in a two classes of models. In the first case the conformal invariance of
the Maxwell field was broken by a non-minimal coupling of the form $RA^{2}$,
this gives a non-zero time-dependent mass to the photon. In the second case
the conformal invariance is violated because of the time-dependent coupling
constant, $I(t)F^{\mu \nu }F_{\mu \nu }$, where $I(t)=I(\phi (t),...)$ is a
general function of non-trivial background fields and $\phi $ can be for
instance inflaton or dilaton.

In principle it looks like inflation can strongly amplify the vacuum quantum
fluctuations and therefore can lead to sizable magnetic fields. However, if
we take into account the back reaction of the electromagnetic field
and require that inflation lasts at least 75 e-folds, the strength of the
primordial field cannot exceed $10^{-32}G$ on $Mpc$ scales and it is not
clear whether such a small field can work as a seed for a possible dynamo
mechanism. 

Only in the strong coupling case, $I(t)F^{\mu \nu }F_{\mu \nu }$, where $%
I=I_{f}\left( a/a_{f}\right) ^{n}$ and $n\simeq 2.2$, the amplitude can
reach the interesting value of $10^{-7}G$ today. However, this case
corresponds to the situation when the effective coupling constant is
extremely large at the beginning of inflation and becomes of the order of
one at the end of inflation and hence the theory is not trustable.\newline
We conclude therefore that the models considered above are not efficient in producing
primordial magnetic fields during inflation and, even if the galactic dynamo was effective, 
the field produced seems to be too small to play the role of a seed 
for this mechanism.

\vspace{15pt}

\textbf{Acknowledgments}: It is a pleasure to thank G. Gabadadze, G. Dvali and A. Gruzinov for useful discussions. V. D. is grateful to L. Vanzo for stimulating discussions. This work was supported by TRR 33 ``The Dark Universe'' and the Cluster of Excellence EXC 153 ``Origin and Structure of the Universe''. V. D. was supported during the first stage of the work by Istituto Nazionale di Fisica Nucleare and University of Trento (Italy).

\end{document}